\documentstyle[12pt]{article}

\newtheorem{theorem}{Theorem}

\newtheorem{lemma}{Lemma}
\newcommand{\rf}[1]{(\ref{eq:#1})}
\newcommand{\be}{\begin{equation}}
\newcommand{\te}{\end{equation}}

\newcommand{\Y}{{\cal Y}}
\newcommand{\R}{{\cal R}}
\newcommand{\N}{{\cal N}}
\newcommand{\Z}{{\cal Z}}

\newcommand{\G}{{\cal G}}

\newcommand{\M}{{\cal M}}
\newcommand{\ud}{\underline}
\newcommand{\I}{ {\bf I}}

\begin{document}

\title{A Fundamental Theorem of Space-times}
\author{E. M.  Monte\\
Universidade de Bras\'{\i}lia, Departamento de Matem\'{a}tica\\
Bras\'{\i}lia, DF. 70910-900 and\\
Universidade Federal  da Para\'{\i}ba, Departamento de Matem\'atica\\
Campina Grande, Pb. 58109-000\\
and\\
M. D. Maia\thanks{E-Mail: maia@mat.unb.br}\\
Universidade de Bras\'{\i}lia, Departamento de Matem\'{a}tica\\
Bras\'{\i}lia, DF. 70910-900}

\maketitle

\begin{abstract}
The fundamental theorem  of  submanifolds is
adapted  to space-times. It is shown that the  integrability
conditions for the existence  of submanifolds of a pseudo-Euclidean  space
contain the Einstein and Yang-Mills  equations.

\end{abstract}
preprint UnB/MAT/FM-M495\\
pacs {02.40 =m, 04.20.Cv, 11.10.Kk}

\section{Introduction}
The fundamental  theorem for  submanifolds  states that a d-dimensional
manifold may be  isometrically  and locally embedded in a D-dimensional
space $\M_{D}$ if and only if its metric, twisting vector and the
extrinsic curvature  satisfy the Gauss, Codazzi
and Ricci equations. Although the  application of this theorem  to  space-times
has been  known for a long time, its physical applications   remain largely
speculative \cite{Fronsdal:1,Joseph,Ne'emman,Friedman},

In a previous paper \cite{MM:1} (hereafter  refereed to simply as {\bf I}),
it was shown that  a given  space-time may  have different embeddings with the
same dimension.  Therefore, we would have a conflicting situation involving
a well known  mathematical  property of  the space-times and
some basic  physical  principles  that  hold  in the space-times.  For
example,  a perfectly   causal  space-time  may exhibit  closed  time-like
geodesics  when an improper local embedding is considered \cite{Fronsdal:2}.
To fix  this signature ambiguity we  need  to
have a  true 1:1  embedding with the  least number of embedding dimensions.

Another property shown in {\bf I}, is that the torsion or twisting
vector (To
avoid confusion with  torsion in Einstein-Cartan geometry we opted to  use the
designation of "twisting vector" instead of  torsion vector)
induces  a  connection  in space-time (called the  twisting connection)  which
transforms  like   a gauge potential
under a  subgroup of the  embedding symmetry. The
curvature associated with such potential is related to the extrinsic curvature
by Ricci's equations. This means that  two  of the  fundamental  equations of
submanifolds, Gauss'  and Ricci's  equations,  have similar
structures, namely they describe the curvature
tensors of the  Levi-Civita  and the  twisting   connections respectively
as  algebraic functions of the  extrinsic curvatur$e^{1}$:
\begin{eqnarray}
R_{ijkl} &= &  b_{i[k}b_{j]l} -  b_{i[l}b_{j]k}\label{eq:G},
\vspace{3mm}\\
F_{ij}\;\;\; &= & 2g^{mn}b_{n[i}b_{j]m} \label{eq:R} ,
\end{eqnarray}
On the other hand,  Codazzi's equations  has  a  different  meaning and it read
as
\be
D_{[j}b_{k]i} =  \;\;\; 0 \label{eq:C}
\te
where  $D_{i}$ is  the covariant derivative associated with the  twisting
connection $A_{i}$.

In the  present note  we show that the  Einstein and Yang-Mills  equations
constitute a  subset
of \rf{G}, \rf{R}, with the extrinsic curvature playing the role of a  source
field, while \rf{C}  has the  character of  a  primary constraint.
Finally,  we  eliminate  the
extrinsic curvature  among \rf{G} and \rf{R},  obtaining a four-dimensional
foliation of space-times where each leaf is a solution of the
Einstein-Yang-Mills equations, with a possible hint to
a  ``geometrodynamical unification  of the  fundamental interactions''.

\section{Involutive Connections}
Let the signature of the embedding space  be  $(p,q)$.
Since the tangent space of the embedded space-time is  Minkowski's space-time,
(with signature $(3,1)$), it follows that the  orthogonal space  has  an
isometry group  $SO(p-3,q-1)$ which is  a proper subgroup of the
homogeneous embedding symmetry  $SO(p,q)$.

As in  \I, let    $L^{AB}$ be the  Lie algebra generators of
$SO(p-3,q-1)$. To this Lie algebra  and to  the  metric  $g_{AB}$  there
is  an associated  Clifford algebra  with generators  $E^{A}$ and  identity
$E^{0}=1$  such that
\be
E^{(A}E^{B)}=g^{AB}E^{0},\;\; \gamma\, L^{AB}=[E^{A},E^{B}], \;\;
\nabla_{i}E^{A}=0,\; \mbox{and}\; \nabla_{i}L^{AB}=0 \label{eq:alg}
\te
where $\gamma$ is  a  normalization  factor to be  adjusted latter. Equation
\rf{alg} suggests that the  Clifford algebra plays the role of a  square root
of the  Lie algebra.

A given distribution $\{\xi_{i}\}$ on a manifold
$\M_{D}$, is said to be Involutive  if $[ \xi_{i} , \xi_{j}]
=\phi^{k}_{ij}\;\xi_{k}$, where $\phi^{k}_{ij}$ are some functions in $\M_{D}$.
 The
local Frobenius theorem
states that an involutive distribution $\{\xi_{i}\}$ is  also
integrable \cite{Boothby:1}. In particular, an involutive distribution of
independent vector fields integrate as  a submanifold of $\M_{D}$
\cite{Sternberg:1}, \cite{Jacobowitz:1}.

Now consider  a distribution associated with  a  connection  $\Gamma_{i}$,
defined
in $\M_{D}$ by the respective  covariant derivative
${\cal D}_{i}$.  We  say that the connection  $\Gamma_{i}$ is involutive if for
a set of independent vector fields $\{X_{k}\}$ we have
\be
[{\cal D}_{i}, {\cal D}_{j}] X_{k} = \phi^{l}_{ijk}X_{l}  \label{eq:FROBENIUS}
\te
where   $\phi^{l}_{ijk}$ are functions on  $\M_{D}$.
In general $\Gamma_{i}$ has a non zero curvature  given by the
the Lie  product $[{\cal D}_{i}, {\cal D}_{j}] $. Therefore, the Frobenius
theorem for  connections may be regarded
as an statement on the curvature of  that connection.  Let us
apply this concept  to  two particular cases: The  Levi-Civita connection of
$\M_{D}$ and the  twisting connection $ A_{i}$ described in {\bf I}.

\begin{lemma}
The  Levi-Civita and the twisting connections are integrabel only if
the Codazzi equation \rf{C} holds.
\end{lemma}
Let us start with the two basic equations for  the  existence
of  a  four-dimensional embedded space-time in a D-dimensional flat space
$M_{D}(p,q)$
\begin{eqnarray}
\nabla_{j}\Y^{\mu}_{;i} & = & -g^{MN}b_{ijM}\N^{\mu}_{N} \label{eq:GW1},\\
\nabla_{k}\N^{\mu}_{A} & = & -g^{mn}b_{kmA} \Y^{\mu}_{,n}
+g^{MN}A_{kAM}\N^{\mu}_{N}
\label{eq:GW2}
\end{eqnarray}
Taking the covariant derivative of \rf{GW1} and after  exchanging  the indices
and  subtracting  we obtain
\[
[\nabla_{k},\nabla_{j}]\Y^{\mu}_{,i}
=-2g^{MN}b_{i[jM;k]}\N^{\mu}_{N}+2 g^{MN}b_{i[jM}\N^{\mu}_{N;k]}
\]
or,  after using \rf{GW2},
\be
[\nabla_{k},\nabla_{j}]\Y^{\mu}_{,i}=
2g^{MN}\left (\nabla_{[k}b_{j]iM}-g^{PQ}b_{i[jQ}A_{k]PM}\right ) \N^{\mu}_{N}
-2g^{mn}g^{MN}b_{i[jM}b_{k]mN} \Y^{\mu}_{,n}.\label{eq:NN}
\te
Introducing the notation
\be
D_{Mk}^{Q}=\delta_{M}^{Q}\nabla_{k}-g^{PQ}A_{kPM} \label{eq:D},
\te
the expression  \rf{NN} may be written  as
\be
[\nabla_{k},\nabla_{j}]\Y^{\mu}_{,k}  =2g^{MN}D_{M[k}^{Q}b_{j]iQ}\N^{\mu}_{N}
-2 g^{mn}g^{MN} b_{i[jM}b_{k]mN} \Y^{\mu}_{,n}. \label{eq:NN1}
\te
As we see, $\nabla_{i}$ is not involutive
bacause the presence of the term in $\N^{\mu}$ in \rf{NN1}. Therefore,  if we
impose the
condition
\be
D_{M[k}^{Q}b_{j]iQ}=0 \label{eq:CODAZZI},
\te
we obtain
\be
[\nabla_{k},\nabla_{j}]\Y^{\mu}_{,i}=  \phi^{n}_{ijk}\Y^{\mu}_{,n},
\label{eq:inv}
\te
with
\[
\phi^{n}_{ijk}= 2 g^{mn}g^{MN} b_{i[jM}b_{k]mN}.
\]
In this case, the expression of  Riemann tensor is
\(
[\nabla_{k},\nabla_{j}]\Y^{\mu}_{;i}=g^{mn}R_{ijkm}\Y^{\mu}_{,n}
\)
and comparing with  \rf{inv}, we obtain
\[
R_{ijkm}\Y^{\mu}_{,n}=-2g^{MN}g^{mn}b_{i[jM}b_{k]mN}\Y^{\mu}_{,n}.
\]
After  replacing
$g^{MN}=E^{(M}E^{N)}$, we obtain Gauss' equation  \rf{G}.

To see that  equation \rf{CODAZZI} is the same as \rf{C}
recall that the  torsion vector  is   a  Lie
algebra defined object $ A_{i}= A_{iAB}L^{AB}$ where  $L^{AB}$ are the  Lie
algebra generators of group  $SO(p-1,q-3)$. The  associated  covariant
derivative  is then given by   $D_{i}=\nabla_{i} -A_{i}$ so that
\[
D_{i}X=\nabla_{i}X -[A_{i},X]
\]
where  $[A_{i},X]$ is defined in the  same algebra as that of  $X$.
Therefore, for   $b_{ij}=b_{ijA}E^{A}$, we have
\[
D_{k} b_{ij}=b_{ijA;k}E^{A} -A_{kCB}b_{ijA}\frac{8}{\gamma}g^{A[B}E^{c]}
\]
or,  taking  $\gamma=8$  in \rf{alg}, it follows that
\[
D_{k}b_{ij}=   (\delta^{A}_{C}\nabla_{k} -g^{AB}A_{kCB})b_{ijA}E^{C}=
D_{Ck}^{A}b_{ijA}E^{C}
\]
therefore,
\be
D_{[k}b_{i]j}=D^{Q}_{A[k}b_{i]jQ}E^{A}.
\te
which, except by a  sign is the left hand side of \rf{CODAZZI}.

Next consider the  covariant derivative $D_{i}$.
Since  $\N^{\mu} = \N^{\mu}_{A}E^{A}$ belongs to the  Clifford  algebra, then
it follows that
\be
D_{i}\N^{\mu}=  \N^{\mu}_{A;i} E^{A}
-A_{iMN}\N^{\mu}_{A}g^{A[N}E^{M]}=D_{jM}^{A}\N^{\mu}_{A} E^{M} \label{eq:DN}
\te
Using Weingarten's  expression in \rf{GW2} with the notation \rf{D},
\be
D_{jA}^{N} \N^{\mu}_{N} =-g^{mn}b_{jmA}\Y^{\mu}_{,n}
\te
so that  \rf{DN} becomes
\be
D_{i}\N^{\mu} = -g^{mn}b_{imM}\Y^{\mu}_{,n}E^{M}=-g^{mn}b_{im}\Y^{\mu}_{,n}
\label{eq:W2}
\te
and the  second  covariant  derivative of $\N^{\mu}$  is
\[
D_{i}D_{j} \N^{\mu}= -g^{mn}b_{jm}\Y_{in}-g^{mn}\Y^{\mu}_{,n}D_{i}b_{jm}
\]
Therefore,
\[
D_{i}\left( D_{j} \N^{\mu}\right)-D_{j}\left( D_{i}\N^{\mu}\right)=
[D_{i},D_{j}]\N^{\mu}=
2g^{mn}\Y^{\mu}_{,n[j} b_{i]m}+2g^{mn} \Y^{\mu}_{,n} D_{[j}b_{i]m}
\]
Or, after using Gauss' equation \rf{GW1},
\[
[D_{i},D_{j} ] \N^{\mu}=
2g^{mn}g^{MN}b_{m[iM}b_{j]n}\N^{\mu}_{N} -2 g^{mn}D_{[j}b_{i]m}\Y^{\mu}_{,n}
\]
Now, imposing Codazzi's equation  \rf{C} we obtain
\begin{equation}
[D_{i},D_{j}]\N^{\mu}= 2g^{mn}g^{MN}b_{m[iM}b_{j]n} \N^{\mu}_{N}
\end{equation}
This last expression may also be written in terms of  components as
\[
([D_{i},D_{j}]\N^{\mu}_{A})E^{A}=2g^{mn}g^{MN}b_{m[iA}b_{j]nM}\N^{\mu}_{N}E^{A}
\]
so that
\be
[D_{i},D_{j}]\N^{\mu}_{A}=  \phi^{N}_{ijA} \N^{\mu}_{N}  \label{eq:invol}
\te
where  we have denoted
\be
\phi^{N}_{ijA}=2g^{mn}g^{MN}b_{m[iA}b_{j]nM}
\te
Therefore,  expression \rf{invol} says that  $D_{i}$ is involutive only after
we apply  \rf{C}.

\hfill{\rule{2mm}{2mm}}

Since we cannot  cancel  $\N^{\mu}$ in \rf{invol}  the explicit expression
for  $[D_{i},D_{j}]$  is more conveniently derived from the definition:
$[D_{i},D_{j}] =-(\nabla_{i}A_{j}-\nabla_{j}A_{i}-[A_{i},A_{j}])$.
In  terms of components this reads as
\[
[D_{i},D_{j}]_{AB}L^{AB} =-(\nabla_{i}A_{jAB}-\nabla_{j}A_{iAB})L^{AB}
-A_{iMN},A_{jPQ}f_{AB}^{MNPQ}L^{AB}
\]
where we have used  the   structure constants of  the  Lie algebra of
$SO(p-3,q-1)$ given in {\bf I}:
\[
[L^{MN},L^{PQ}]=f_{AB}^{MNPQ}L^{AB}=2
\delta_{A}^{[N}g^{m][P}\delta_{B}^{Q]}L^{AM}
\]
Therefore, we obtain Ricci's  equation \rf{R}
\[
F_{ij}=[D_{i},D_{j}]= 2g^{mn}b_{m[iA}b_{j]nB}L^{AB}=
\frac{1}{2}g^{mn}b_{m[i}b_{j]n}
\]

\section{The Fundamental Theorem for Space-times}
The fact that  the twisting vector
$A_{i}$ transforms  as  a  gauge field
suggests that  a Yang-Mills  theory of geometrical nature is
contained in the embedding of the space-time.
\begin{lemma}
The torsion vector of  a  space-time is a  Yang-Mills field with respect to
the group  SO(p-1,q-3), defined by the equations
\begin{eqnarray}
D^{i}F_{ij}  & = & {\bf j}_{j}^{geom.}, \label{eq:YM}\\
D^{i}F^{jk}& + & D^{k}F^{ij}+D^{j}F^{ki}=0, \label{eq:Jacobi}
\end{eqnarray}
where
\begin{equation}
{\bf j}_{j}^{geom.}=\frac{1}{2}g^{ik}g^{mn}\left
([b_{mi},D_{k}b_{jn}]-[b_{mj},D_{k}b_{in}] \right ) .
\label{eq:SOU}
\end{equation}
\end{lemma}
In fact,  since $ [D^{i}, [D^{j},D^{k}]]
= D^{i}F^{jk} $, the
second equation is a direct consequence of Jacobi's identity:
\[
[D_{i}, [D_{j}, D_{k}] ]+[D_{k},[D_{i},D_{j}]]+[D_{j},[D_{k},D_{i}] ]=0
\]
On the other hand,  taking the covariant divergence of  $F_{ij}$  and using
$D_{k}g^{mn}=0$,
we find that
\be
D^{i}F_{ij} =g^{ik}D_{k}\left ( g^{mn}b_{m[i}b_{j]n}  \right )
=\frac{1}{2}g^{ik}g^{mn}
\left ( [b_{mi},D_{k}b_{jn} ]-[b_{mj},D_{k}b_{in}] \right ).
\te
Therefore,  with the definition   \rf{SOU}, we obtain   \rf{YM}.
\hfill{\rule{2mm}{2mm}}

Notice  that $D_{i}$ contains  $A_{i}$   and therefore this  connection cannot
be eliminated  from ${\bf j}_{j}$, even  taking into account Codazzi's
equation. Consequently,  the solution of equation
\rf{SOU}  in general  depends on an  integration over compact surfaces,
so that  ${\bf  j}_{i}^{geom.}$  may be associated  to  a topological charge.
The  covariant derivative of the first equation gives
\[
D^{j}D^{i}F_{ij} = D^{j}{\bf j}_{j}^{geom.}= 0
\]
which says that   ${\bf j}_{j}^{geom.}$ is  a  conserved current.
In the following we  will see that  the   current ${\bf j}_{i}^{geom.}$ is
related  to  the  source  of the gravitational field.

The   mean curvature  $h$ and   $k$  is the scalar
extrinsic curvature of the space-time,  are respectively  given by
\be
\begin{array}{l}
h^{2}=g^{AB}h_{A}h_{B}, \;\; \;\;\;\; h_{A}=g^{ij}b_{ijA}.
\vspace{3mm}\\
k^{2}=g^{AB}b_{miA}b^{mi}_{\;\; B}=b_{miA}b^{miA},
\end{array}
\te
\begin{lemma}
The  metric of the embedded space-time is  necessarily  a solution of
\be
G_{ij}=  R_{ij}-\frac{1}{2}Rg_{ij} ={\bf t}_{ij}^{geom.} \label{eq:EG}
\te
where
\begin{equation}
{\bf t}_{ij}^{geom.}=b_{imA}b_{j}^{\;mA}
-h_{A}b_{ij}^{\;\; A} -\frac{1}{2}(k^{2}-h^{2})g_{ij}  \label{eq:TAU}
\end{equation}

\end{lemma}
This  is a direct consequence of  the contractions of  Gauss equation
\rf{G}:
\be
{R}_{jk}=g^{il}R_{ijkl}=2g^{MN}g^{mn}b_{j[mM}b_{n]kN}\label{eq:X}
\te
and
\be
{R}= g^{jk}R_{jk}=2g^{MN}g^{jm}g^{kn}b_{j[mM}b_{n]kN}=h^{2} -k^{2}
\label{eq:SCL}
\te
The expression \rf{EG} follows immediatly.

\hfill{\rule{2mm}{2mm}

Since by  hypothesis our embedded manifold is a
solution of Einstein's equations  for a given  source,   $G_{ij}=8\pi G\, {\bf
t}_{ij}^{matter}$, then, \rf{EG}  is equivalent  to
\begin{equation}
{\bf t}_{ij}^{geom.}= 8 \pi G\; {\bf t}_{ij}^{matter}
\label{eq:TT}
\end{equation}
which is  an algebraic (non-differentiable)  equation relating  $b_{ij}$ to
the matter fields.

\vspace{0.5cm}

As  an example  consider Scwarzschild's space-time with  the
six dimensional embeddings  given Fronsdal and in {\bf I}. The Ricci flat
condition implies that ${\bf  t}_{ij}^{geom.}=0$, or $h=k$.
In this case the gauge group of the twisting connection  is
$SO(2)$ so that $A_{i}$ is  an electromagnetic  field of  geometrical
nature.  In the case of  the  Kasner embedding  also given in {\bf I},
we would have a gauge group  $SO(1,1)$.

Another  simple example is given by a  five dimensional flat embedding
space. In this case the  twisting vector  does not exist and a general
expression of  $b_{ij}$ is given by \cite{Szk}:
\[
b_{ij}= \lambda\, g_{ij} +(4\lambda-h)u_{i}u_{j}
\]
where  $\lambda$ depends on the distrubution of matter in space-time  and
$g^{ij}u_{i}u_{j}=-1$. This  severely  restricts the type of  matter allowed in
the space-time.  For  example, in a  space-time filled with  dust,
$\; {\bf t}_{ij}^{matter} = \rho u_{i}u_{j}$, where $\rho$
is the matter density,  In this case \rf{TT}    gives
(taking $8\pi G=1$)
\[
g^{mn}b_{im}b_{jn} -hb_{ij} -\frac{1}{2} (k^{2}-h^{2})g_{ij}=-\rho
u_{i}u_{j}
\]
so that  $ k^{2}-h^{2}= -\rho$, which is satisfied if
$b_{ij}= \left( h-\frac{\rho}{h}\right)g_{ij}$ and
$h^{2}= 4\rho /3 $, so that  the extrinsic curvature  is  in direct proportion
to  the square root of the  density:
\[
b_{ij}  = \frac{1}{2}\sqrt{\frac{\rho}{3}}\,\; g_{ij}
\]

\hfill{\rule{2mm}{2mm}

\begin{theorem}[Fundamental Theorem].\\
Given a space-time of general relativity corresponding to a source
${\bf t}_{ij}^{matter}$, then it has a  unique local and isometric embedding
in a pseudo-Euclidean space  $M_{D}(p,q)$,  with the least number of
dimensions, provided there is  a  Yang-Mills
gauge field $A_{i}$, ($D>5)$,  with gauge group $SO(p-3,q-1)$ satisfying
Ricci's
equation \rf{R} and   a  symmetric  second order tensor $ b_{ij}$  satisfying
Codazzi's equation \rf{C} such that
\[
{\bf t}_{ij}^{geom.}={\bf t}_{ij}^{matter}  \;\;\;{and}\;\; \;
{\bf j}^{geom.}_{i}={\bf  j}^{matter}_{i}
\]
\end{theorem}
In fact, any Riemannian or pseudo-Riemannian  manifold  can in principle be
embedded in some pseudo-Euclidean space and in particular  this applies to
space-times.

{}From  the results of  {\bf I}  we  have seen
that if the  number of embedding dimensions is the smallest  possible,
we  have  an well defined embedding without  ambiguities in the signature.
In this  case,  the twisting vector  transforms  as  a Yang-Mills
fields  with group $SO(p-3,q-1)$ and lemma 1 says that it obeys a
Yang-Mills like  equation with current  ${\bf j}_{i}$ given by \rf{SOU}. On the
other hand, the metric $g_{ij}$ is by definition  the  gravitational
field and  from  lemma 2  its source is  related to  $b_{ij}$ by
\rf{TT}.

Reciprocally, suppose  we have  are given a real tensor  $g_{ij}$, a
$SO(p-3,q-1)$-Lie algebra vector $A_{i}$  and  a Clifford algebra tensor
$b_{ij}$ such that
\be
\begin{array}{l}
 G_{ij}={\bf t}_{ij}^{matter}\vspace{3mm}\\
 D^{i}F_{ij}={\bf j}_{j}^{matter}
\end{array}
\te
With the solutions of these equations we  may  write the  respective curvature
tensors $R_{ijkl}$  and $F_{ij}$. Replacing these curvatures in  equations
\rf{G},\rf{C} and \rf{R},  we obtain the system of equations to determine
$b_{ij}$:
\[
 D_{[j}b_{k]i} = 0,\;\;
b_{[ik}b_{lj]}-b_{[jk}b_{li]}=R_{ijkl},\;\; \mbox{and}\;\;
g^{mn}b_{n[i}b_{j]m}=\frac{1}{2} F_{ij}
\]
Therefore, we obtain a complete set of
quantities  $g_{ij}$, $A_{iAB}$ and  $b_{ijA}$  satisfying
the integrability conditions
for the  embedding of   the  space-time in $M_{D}(p,q)$.

\hfill{\rule{2mm}{2mm}

\section{Space-time Foliations}
 In this section we explore the possibility that the  $b_{ijA}$ field may be
eliminated between the equations \rf{YM} and \rf{EG}.

Consider the hypothetical  situation where
physics is not necessarily confined to the four-dimensional hypersurface of
$\M_{D}$. That is,  by some high energy process,
some particles which would be otherwise
trapped  in the  space-time surface will now escape this  constraint and move
along the extra dimensions. This  is similar to  the model proposed by
Rubakov \& Shaposhnikov \cite{Rubakov:1}:
Since our embedding space is flat  a  (classical) particle  in the vicinity of
space-time may be described by the coordinates
\[
\Z^{\alpha}(x^{i}, x^{A})=\Z^{\alpha}(x^{i}, 0)+x^{A}\N^{\mu}_{A}
\]
The result is a (multiparameter)  foliation of  $M_{D}$  with parameters
$x^{A}$,  where each leaf has  the metric (the same  used in \I)
\be
\gamma_{\alpha\beta}
=\Z^{\mu}_{\alpha}\Z^{\nu}_{\beta}\G_{\mu\nu}
= \left
(\begin{array}{cc}
              \tilde{g}_{ij} + g^{MN}A_{iM}A_{jN}  &  \; A_{iM}\\
			  A_{jN}               &   \; g_{AB}
\end{array}
\right
)\label{eq:GAMMA}
\te
where we have denoted
\be
\begin{array}{lll}
\tilde{g}_{ij} = g^{mn}(g_{im} -x^{A}b_{imA} )(g_{jn}-x^{B}b_{jnB})
\vspace{3mm}\\
A_{iM}= x^{A}A_{iMA}
\vspace{2mm}\\
g_{AB}=\epsilon_{A} \delta_{AB},\;\;\; \epsilon_{A}=\pm 1
\end{array}
\te
Since  $M_{D}(p,q)$ is  flat,
the  Einstein-Hilbert Lagrangian for $\gamma$ gives (after using an analogy
with the Kaluza-Klein metric ansatz)
\be
\R(\gamma)\sqrt{\gamma}  = \tilde{R}(\tilde{g})\sqrt{\tilde{g}}
+\frac{1}{4}tr F^{ij}F_{ij}  =0
\te
The classical  equations
obtained from this Lagrangian are the  Einstein-Yang-Mills equations  for the
for the metric $\tilde{g}_{ij}$
\be
\tilde{R}_{ij} -\frac{1}{2}\tilde{R} \tilde{g}_{ij} = {\bf t}_{ij}(F)
\label{eq:EYM}
\te
where in the right hand side  we have the energy momentum tensor of  the
$A_{i}$ field and all contractions  are  made with  the  metric
$\tilde{g}_{ij}$
of the  4-dimensional  solutions.

\hfill{\rule{2mm}{2mm}

These results  appear to indicate that when properly defined, the embedding
is not only  compatible  with four  dimensional physics but also
that it lies  at the root  of the  fundamental equations of  physics in
space-time.  We were originally motivated by a possible extension of
\cite{Violette}, to determine if  Einstein and Yang-Mills equations
would be sufficient to deteremine the  embedding of the space-time,
As  we have seen, the Codazzi  equation \rf{C} plays an essential role  in the
process and   it cannot be  dispensed with.
After this equation  is imposed,  Gauss and Ricci's equations
become simply conceptually  equivalent as expressions of the  respective
curvatures in terms of the  space-time  extrinsic curvature.
Equations \rf{YM}, \rf{Jacobi} are formally identical to the Yang-Mills
equations relative to the  gauge group  $SO(p-1,q-3)$, whose source  is
derived solely from the  extrinsic  geometry of  space-time.

The last result is  somewhat akin to  geometrodynamics where an electromagnetic
potential of  geometrical nature is  postulated \cite{Wheeler}. As we  see,
such potential  may be  replaced by  the twisting connection $A_{i}$.
Here the  Yang-Mills-Wheeler geons would
be four dimensional compact solutions of \rf{EYM}.
There  are  strong analogies with Kaluza-Klein theory, although the
fundamental  postulates are distinct from that theory. The  expression
\rf{GAMMA}  replaces the
Kaluza-Klein metric ansatz  and  the  four dimensional space-time  is  a
deformation  (a leaf of the foliation) of the original one.

\vspace{2cm}

\begin{center}
{\bf Footnotes}
\end{center}
1) {The notation and
index
convention is the  same as in {\bf I}: Lower case  Latin indices run from 1 to
4 and
capital Latin indices run from 5 to D, where
D is the smallest possible embedding dimension. All Greek indices run from 1 to
D. The indicated antissymetrization applies only to the indices of the same
kind
near the brackets.}

\end{document}